# Report on LEO satellite impacts on ground-based optical astronomy for the Rubin Observatory LSST

Phanindra Kandula, Lee Kelvin, Erfan Nourbakhsh, Daniel Polin, Tom Prince, Meredith Rawls, Adam Snyder, Brianna Smart, Christopher Stubbs, Anthony Tyson, Zeljko Ivezic, Craig Lage, Clare Saunders

# Executive Summary

In August 2025 a workshop was convened to bring together experts to better understand steps that can be taken to mitigate the impact of satellite constellations on astronomical observations. At the time, just over 12,000 operational satellites were in low-Earth orbit (LEO). Although reflected sunlight and emissions all across the electromagnetic spectrum from artificial satellites impact scientific observations and the sky, the workshop focused on reflected sunlight in the wavelength range 330 nm to 1100 nm. This aligns with the Vera C. Rubin Observatory Legacy Survey of Space and Time (LSST) planned imaging observations over the coming decade. Among other conclusions, we affirm previous recommendations that tracked satellite apparent magnitudes should be no brighter than 7th AB mag.

The workshop participants discussed over 30 publications, reports, and presentations, and arrived at the Findings and Recommendations presented here. During the workshop, in addition to affirming many existing recommendations and best practices, the group discovered new issues and devised possible mitigations. These were nearly equally divided between advice to satellite builders and operators and to the observational astronomy community. While the workshop prioritized considerations for LSST, and only one satellite company was directly represented, our hope is that many of the Findings and Recommendations will also apply to other observatories and constellations, and that all satellite companies will continue to engage in dialog with sky observers across the globe.

The ultra-faint universe remains essentially unexplored, both in the static deep sky and for transient events. This is where LSST is unique, and this is where discovery of the unexpected will occur. Most of the interfering satellites or debris are faint and near the limit of detection. The number of satellites in LSST data increases exponentially at the faint end. While only a fraction of LSST's focal plane will be covered by satellite streaks, most science is instead affected by systematic errors introduced by the low surface brightness light. Thus, small satellites or debris from collisions presents a challenge.

Several key findings and recommendations relate to the orbit height of satellites. Satellites orbiting above 700 km pose a larger threat to the safety of all satellites in that region because of the high likelihood of runaway collisions. The recommendation is to fly at much lower orbits where the atmosphere assures deorbit times of years instead of centuries. Satellites in high orbits also can reflect sunlight — a conclusion of earlier workshops. A population of otherwise identical satellites may have a smaller impact at lower orbital altitudes than at higher orbital altitudes for most large survey science investigations. Therefore, there is a need to develop constellation-wide metrics alongside per-satellite limits. It is also necessary to develop new standards to enable sharing of satellite brightness observations, simulations, and experimental test results.

Recommendations are given on measures that satellite builders can take to ensure that the satellite meets brightness requirements for avoidance of interference with LSST and ground-based observatories generally. Satellite operators should endeavor to meet these requirements, through a combination of satellite design, construction, and operations. These are expressed in engineering units for the satellite components such as watts per steradian per square meter, integrated over a wavelength band. Reflectivity of components may be adjusted in order to meet these goals. These recommendations are a function of satellite size, orientation, telescope size, and reflectivity, and wavelength. We note these are optimized to the particular optical wavelengths at which LSST observes, and that minimizing interference in the infrared, thermal, and radio regimes is also important to other scientific facilities.

Reflected light from communications satellites are not the only threat. Recently there have been proposals for thousands of large satellites in LEO which will illuminate the ground at night. This is potentially ruinous for ground-based optical astronomy in general.

We recommend that satellite operators share timely satellite ephemerides publicly. Knowledge of accurate satellite ephemeris data is vital to the suppression of contamination of LSST images. Including uncertainty information in the form of covariance matrices, and using the recommended standard OEM format for the ephemerides, is strongly recommended.

It is essential to consider the full satellite life cycle, and both design and operations, when evaluating darkening mitigations. Projections indicate that as many as 100,000 LEO satellites might be in orbit by the 2030 timeframe. To date, most recommendations and mitigations have focused on when satellites are in their main operational lifetimes, i.e., “on station.” However, in practice, many satellites spend weeks to months in a transitional lower-altitude holding orbit before raising their orbits. In addition, with the introduction of the FCC five-year rule, most satellites are deorbited within five years, and many much sooner than this. To maintain a population of 100,000 satellites with five year lifetimes, around 50 will be launched and deorbited per day. We therefore recommend coordinated daytime de-orbits in a short time over the Pacific Ocean, as well as waiting the full five years whenever practicable, to minimize interference with ground-based astronomy.

We recommend the construction of a calibrated comprehensive database of LSST streaks to ultra faint levels. Most LSST discoveries will occur at the faintest levels of surface brightness. This is where LSST is unique: the discovery of the unexpected. Both static deep sky and time-domain data will contain induced systematic errors. We therefore suggest using a novel automated faint streak detection and analysis pipeline that populates a comprehensive database of known affected images and catalogs.

Finally, we recommend support for coordinated satellite operational and ground-based observational experiments. Conducting experiments with varying satellite orientation and internal configuration (such as pointing of solar arrays and antennas), while meeting operational constraints, especially while over-flying observatories, will be helpful in furthering our collective understanding of how to minimize impacts on astronomy.

# Introduction and Scope

This report is the product of a three day NSF-sponsored workshop at UC Davis which explored measures that satellite builders and operators, as well as the Rubin Observatory community, could take to mitigate optical interference with the Rubin Observatory and ground based observatories generally. This report is divided into Findings and Recommendations to each community. For several weeks prior to the workshop, committee members had a chance to contribute and read publications to be discussed. Other groups have studied this, and our report builds on many others from the past five years [e.g., 8,9,10,11,16,17,22,34,36]. All publications and reports that were discussed are listed in the references section. Prior to the workshop, the organizers reached out to invite representatives from four companies that currently operate satellites in LEO: SpaceX, Amazon, Eutelsat, and AST SpaceMobile. Only an engineer from SpaceX was able to attend. It is nevertheless the sincere hope of the authors that the Findings and Recommendations will be broadly applicable to any satellite operator that seeks to collaborate with astronomers to protect dark skies.

Our workshop focused on the technical issues of low-Earth orbit (LEO) satellites in Rubin Observatory Legacy Survey of Space and Time (LSST) images, as well as mitigations. Since LSST is the most sensitive optical sky survey currently in operation, most mitigations should apply to other telescopes as well [36]. As satellite constellations expand to enhance global connectivity, our night skies are evolving, creating new considerations for astronomical observations. The long-awaited decade of LSST discoveries will coincide with a growing presence of LEO satellites in images, highlighting the value of joint efforts to minimize impacts. The workshop brought together experts to quantify and ultimately mitigate impacts to key LSST science goals. We studied simulations of satellite trails, flares, glints, as a function of constellation height, and compared to early LSST data to quantify effects on LSST sensors. A key deliverable of this report is aimed at satellite builders and operators: recommended goals for optical brightness characterization (i.e., how satellite operators should measure brightness), including aggregate impacts.

The expected LSST discovery space is dominated by low surface brightness (LSB) science. Legacy sky surveys have covered bright classes of astronomical objects in great detail. To complement this, LSST is designed to explore LSB science in both the static and time-domain spheres. Consequently, the LSST system is optimized to preserve faint signals, ensuring sensitivity to the low surface brightness regime. LSB science necessarily demands exquisite control of systematics and sky backgrounds [14]. While the bright cores of satellite streaks can be masked, their faint wings extend over arcminutes and are difficult to model or remove. These extended features accumulate in co-added images, bias image variance estimates, and can masquerade as genuine diffuse structures such as tidal streams, intracluster light, or faint galaxy outskirts. Analysis of these faint structures – some of the faintest gravitationally bound structures in the Universe – is essential for furthering our knowledge of how galaxies form and evolve, and in constraining contemporary cosmological models. In the time domain, glints and rotating reflections create irregular signatures that resist simple masking. Thus, contamination is not limited to localized pixel loss, but introduces widespread background bias that directly undermines Rubin's core LSB science goals.

In this report, we address urgent programmatic and scientific needs pertaining to satellite-terrestrial coexistence at optical wavelengths. We aim to advance knowledge of the detailed technical challenges for LSST due to the forecast of 100,000 or more LEO satellites by mid-survey (2030). We study the impact on LSST data, camera, and CCD sensors based on existing lab data and full observation simulations including the effects of defocus of the satellite, and we use early commissioning data from LSST ComCam and LSSTCam for validation.

We also study the effect of changing satellite parameters such as constellation height; satellite dimensions/geometry and orientation; and satellite reflectivity using BRDF (Bidirectional Reflectance Distribution Function) measurements of known components. Constellation height also affects the time to deorbit exponentially due to atmospheric drag.

This report gives concrete recommendations to satellite operators and satellite builders for implementable engineering parameters such as total reflected sunlight power in watts/ster, and satellite-observatory avoidance via attitude adjustments. The report also contains recommendations for mitigations by Rubin and other ground-based optical observatories. We note both intentional and unintentional radio frequency emissions across the electromagnetic spectrum, including radiation at 50–200 MHz, as well as GHz radio emissions, can be harmful to other telescope facilities, but this is beyond the scope of this report.

While protecting science results from large research observatories is important, we also acknowledge the importance of protecting the night sky as a whole, as a source of wonder, a repository of traditional knowledge in cultures worldwide, and as a navigation aide.

In order to protect the night sky, it is imperative to consider the brightness of the whole satellite constellation population as viewed from many different latitudes, seasons, and times of night by naked-eye stargazers.

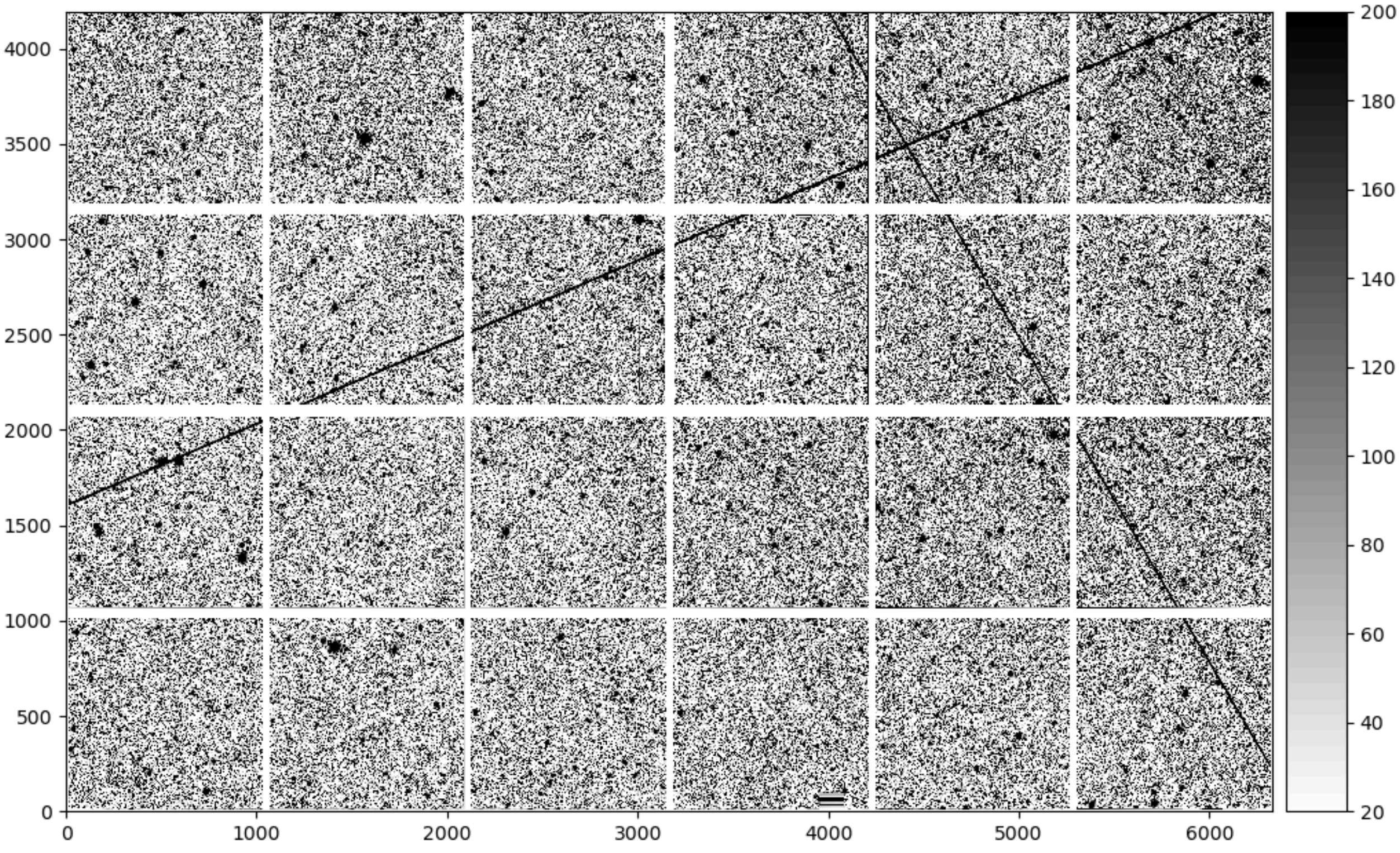

Figure 1. This small portion of a LSST 30-sec 750 nm (*i* band) image taken on July 11, 2025 shows two faint unidentified satellites crossing several of the 189 LSST camera 4Kx4K pixel CCD detectors. The noise is not visible; most of the objects shown are stars in our galaxy. The x- and y-axes are in units of 4 binned pixels (0.8 arcsec), and the intensity bar is in units of electrons/pixel (surface brightness). While the faint streaks fall below the algorithmic detection threshold of the LSST Science Pipelines, so no bogus sources were written to science catalogs, they are clearly present, and may affect other sources that fall in or near the streaks if they are not accurately identified and measured.

## Findings

### 1. Large constellations of LEO satellites are already contaminating most astronomical images, especially near twilight and close to the horizon, and impacts depend strongly on orbital altitude

Reflected sunlight from artificial objects in LEO, including satellites and debris, has a particularly strong impact on astronomical surveys with large mirrors, sensitive cameras, and large fields of view. This precisely describes LSST. Observational surveys that seek to detect potentially hazardous asteroids by making observations near the horizon at twilight, which is one key science goal of LSST, can be affected.

Many darkening mitigation strategies have been explored since large satellite constellations began launching circa 2019. One promising technique, used by SpaceX Starlink and others, is to make the nadir side of the satellite a mirrored surface[1]. This reflects sunlight past a ground based observatory off into space, as shown in the figure below.

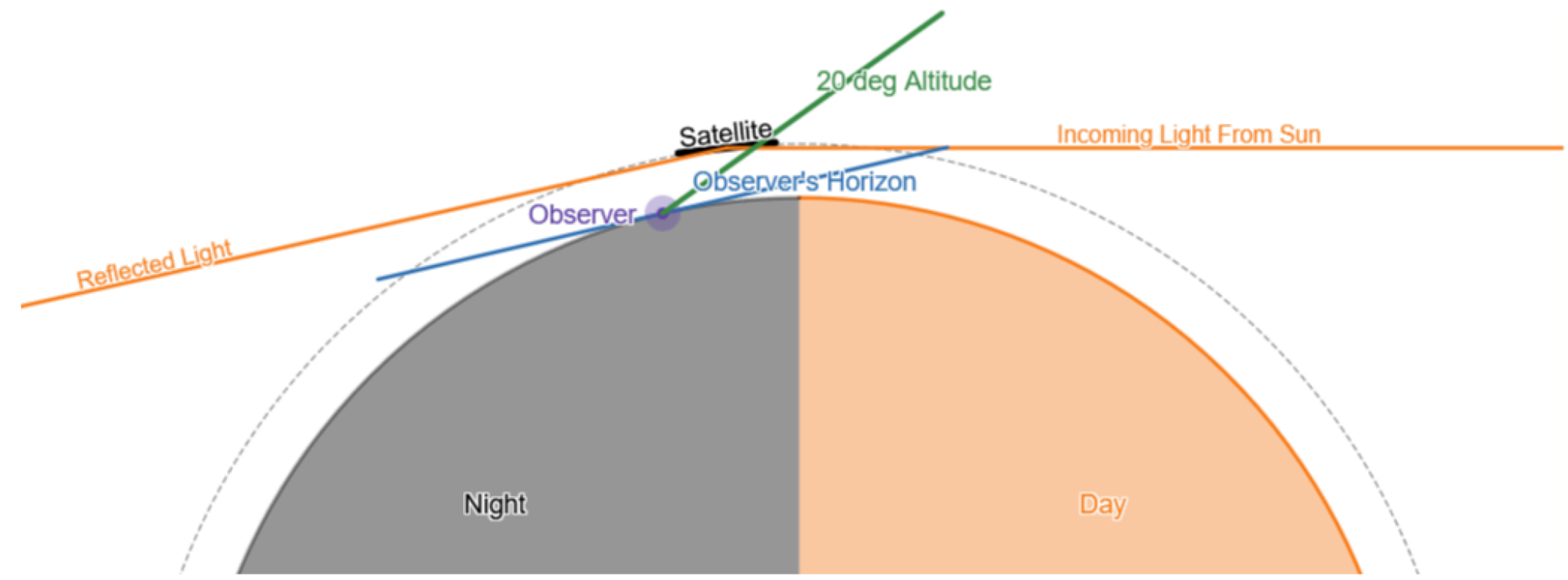

Figure 2. Schematic illustration of how a specular reflective coating on the nadir pointing surface of an object orbiting Earth will, when operating normally, reflect most incident sunlight away from Earth (and thus away from all ground-based observers). Figure from [3].

Of course, most satellites are not perfect spheres or boxes. They have solar panel arrays, antennae, and other components that can be arranged in many geometrical configurations, with a wide array of reflective properties, each oriented at various attitudes. Satellites vary in brightness as a function of time as different surfaces reflect different amounts of sunlight to ground-based observers. Nevertheless, another promising darkening mitigation undertaken by SpaceX Starlink is to orient satellite solar panels so they reflect sunlight away from Earth. It is important to note this orientation is not always possible.

[1] https://www.starlink.com/public-files/BrightnessMitigationBestPracticesSatelliteOperators.pdf

To investigate satellite brightness as a function of many variables, a recent simulation of thousands of Starlink-like satellites was undertaken in [3] in which the reflectivity was determined from repeated photometry of each satellite.

For the general case in which the reflectivities are initially unknown, simulation of LSST operations plus repeated photometry of every satellite in simulated model constellations was carried out. It was found that the Starlink V2 satellites appear brightest at two distinct positions in the sky: when oriented toward the sun at low elevations where light is specularly reflected, and nearly overhead where the satellite is closest to the observer. This is shown for evening twilight in the figure below for satellites at 550 km and 350 km orbit height as the terminator passes overhead.

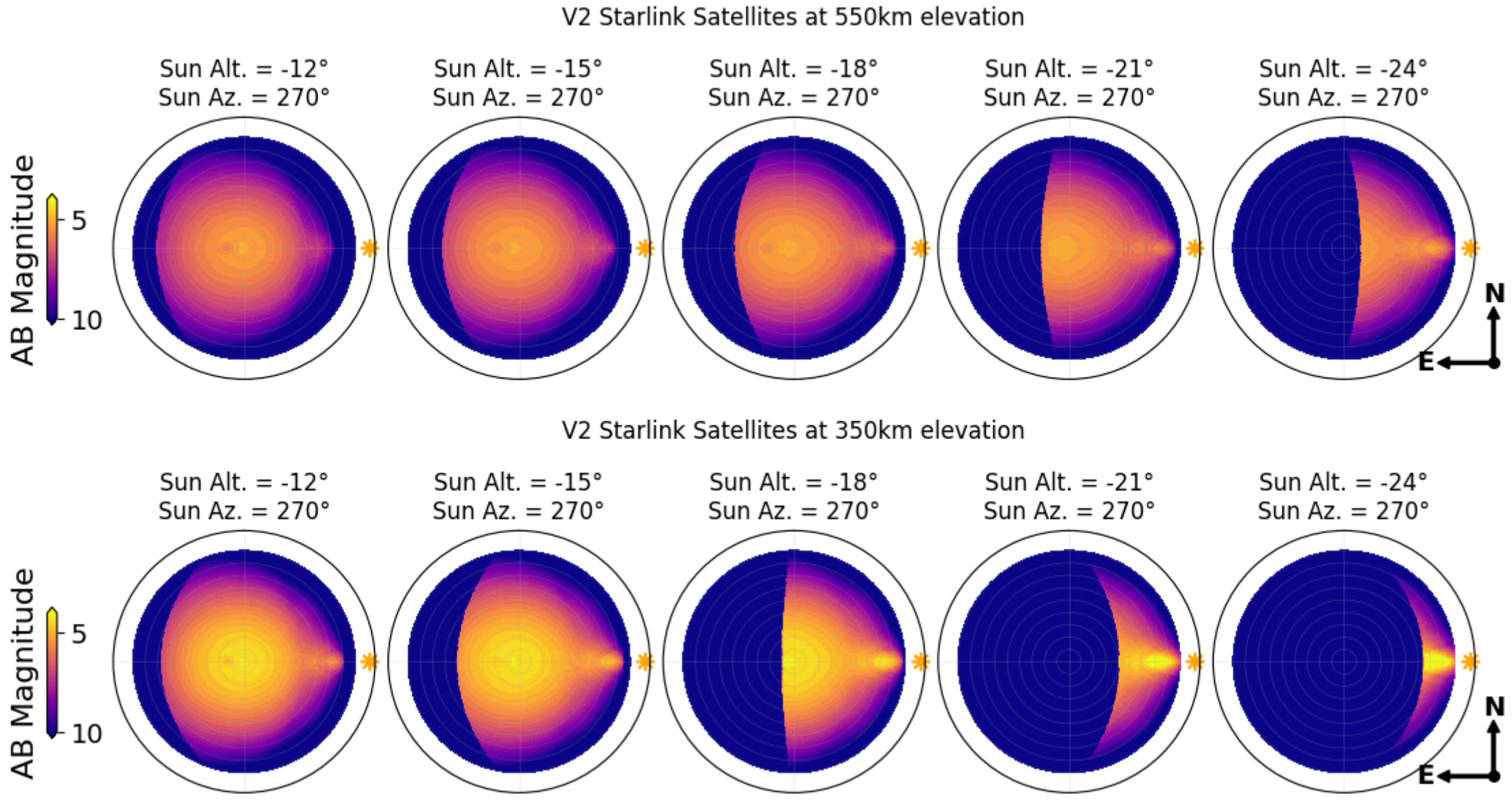

Figure 3. Sky brightness distributions of tens of thousands simulated Starlink V2 satellites in evening twilight over several years. *Top*: 550 km orbital altitude. *Bottom*: 350 km orbital altitude. This figure is similar to those shown in [3]. The satellite's altitude is its elevation above the horizon, and its azimuth is represented by the polar angle. Zenith is in the center of each circle. The Sun position is the orange star below the western horizon. Fewer illuminated satellites are visible sooner when they orbit at 350 km instead of 550 km, but many do have a brighter tracked (effective stationary) magnitude. The bright spot close to the western horizon is due to specular reflections that are well-known to astrophotographers, called cluster flares, a side effect of the mirrored coating that usually reflects light away from Earth. The bright spot near zenith is due to proximity.

If the entire Starlink V2 constellation were lowered to 350 km, [3] finds a 40% reduction in the number of bright satellites entering the LSST focal plane compared to 550 km height. A combination of factors, including lower satellites appearing more out of focus, yield an apparent surface brightness of these satellites for LSST operations only 5% brighter than at 550 km orbit. Overall there is an advantage to LSST for lower altitude satellite constellations.

Of course, details are different for other satellite geometries and component reflectivities, but in general, lower orbits have an advantage for optical survey astronomy, and can also offer advantages for satellite operators. We recognize there are also challenges with lower orbits, such as increased atmospheric drag and susceptibility to space weather events.

## 2. A population of otherwise identical satellites may have a smaller impact at lower orbital altitudes than at higher orbital altitudes for most large survey science investigations

A variety of factors combine to make satellites at lower orbital altitudes, (~350 km) generate a smaller impact on optical astronomy than higher altitudes (~500 km). This is mainly due to reduced streak rate and that the surface brightness of a streak is largely independent of orbital altitude of the constellation [2,3,7]. Due to the earth's shadow, lower altitudes lead to fewer satellites being illuminated over a shorter period of time. Increased orbital speed would decrease the surface brightness somewhat, compensating for the decrease in distance to Earth-based observers. However, studies have shown [3] that the brightest magnitudes of satellites that do still scatter light into telescopes can be somewhat higher.

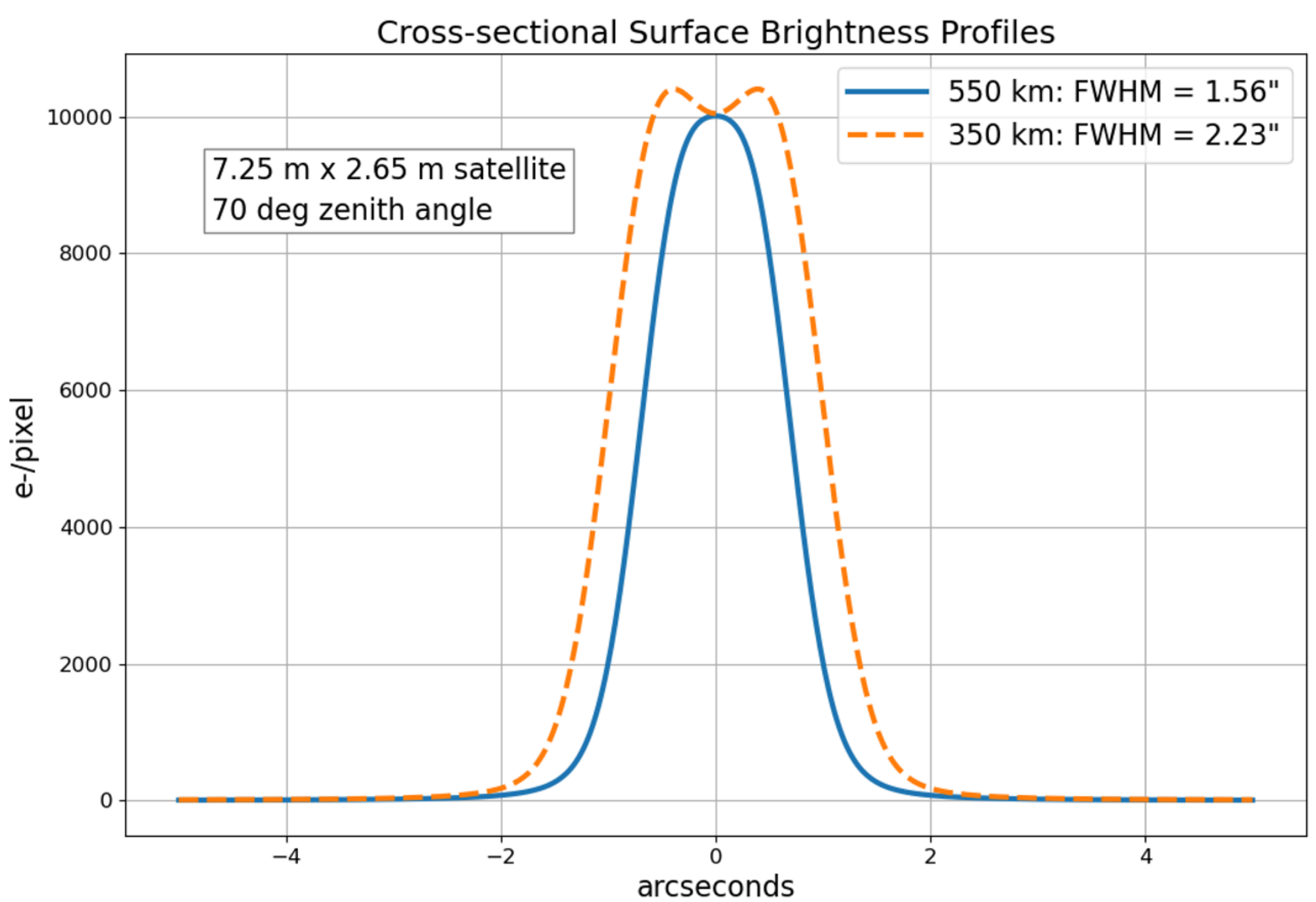

Figure 4. A combination of a satellite's range, angular velocity, and the defocus effect make its resulting streak surface brightness nearly independent of orbital height. In other words, the *blue* (550 km) satellite and an otherwise identical *orange* (350 km) satellite have similar peak surface brightnesses. We note in both cases, the low surface brightness wings extend tens of arcseconds away (orthogonal) from the streak. Given this, we prefer lower altitudes, because fewer satellites are visible at any given time.

The surface brightness of the satellite is inversely proportional to the orbital height due to the $1/r^2$ effect. This is counterbalanced by defocus. The telescope is focused on infinity, whereas the satellite in LEO appears relatively out of focus due to the large LSST mirror.

## 3. It remains challenging to parametrize and model the brightness of satellites without detailed knowledge of component design and operational states

The amount of reflected light intercepted by a telescope has a complex dependence on solar illumination, materials on the satellite, and its orientation and internal configuration. The solar panels and communications antenna can be independently orientated relative to the satellite bus, and this adds complexity to understanding the signatures [1,3,4,13]. Even for relatively simple satellite designs, collaboration with satellite operators who are willing to share BRDF information is usually required to accurately forecast satellite brightnesses.

## 4. It is essential to consider the full satellite life cycle, and both design and operations, when evaluating darkening mitigations

Projections indicate that as many as 100,000 LEO satellites might be in orbit by the 2030 timeframe [3]. To date, most recommendations and mitigations have focused on when satellites are in their main operational lifetimes, i.e., “on station.” However, in practice, many satellites spend weeks to months in a transitional lower-altitude holding orbit before raising their orbits. In addition, with the introduction of the FCC five-year rule[2], all satellites are deorbited within five years of end of life, often much sooner than this. All of these life cycle phases have the potential to be brighter than when a satellite is on station: thousands of such bright satellites by 2030.

The early-stage and late-stage phases of the satellite life cycle deserve particular consideration. During orbit-raising these systems we suspect there is considerable flexibility in orientation while maintaining a low ballistic coefficient. When constellation sizes grow to the expected number, deorbiting will become commonplace and we need to attend to the consequent fireballs in the sky [24,25]. For a population of 100,000 satellites, assuming operators intend to upgrade their satellites every 5 years, as many as 50 will have to be launched and deorbited per day. While techniques have been developed for deorbiting satellites[3], this level of activity by operators will produce a bimodal distribution of LEOsat streak intensity with peaks at high intensity and at very low surface brightness at low intensity. Due to the large numbers, LSST cannot avoid either.

[2] https://www.fcc.gov/document/fcc-adopts-new-5-year-rule-deorbiting-satellites-0
[3] https://www.starlink.com/public-files/Starlink_Approach_to_Satellite_Demisability.pdf

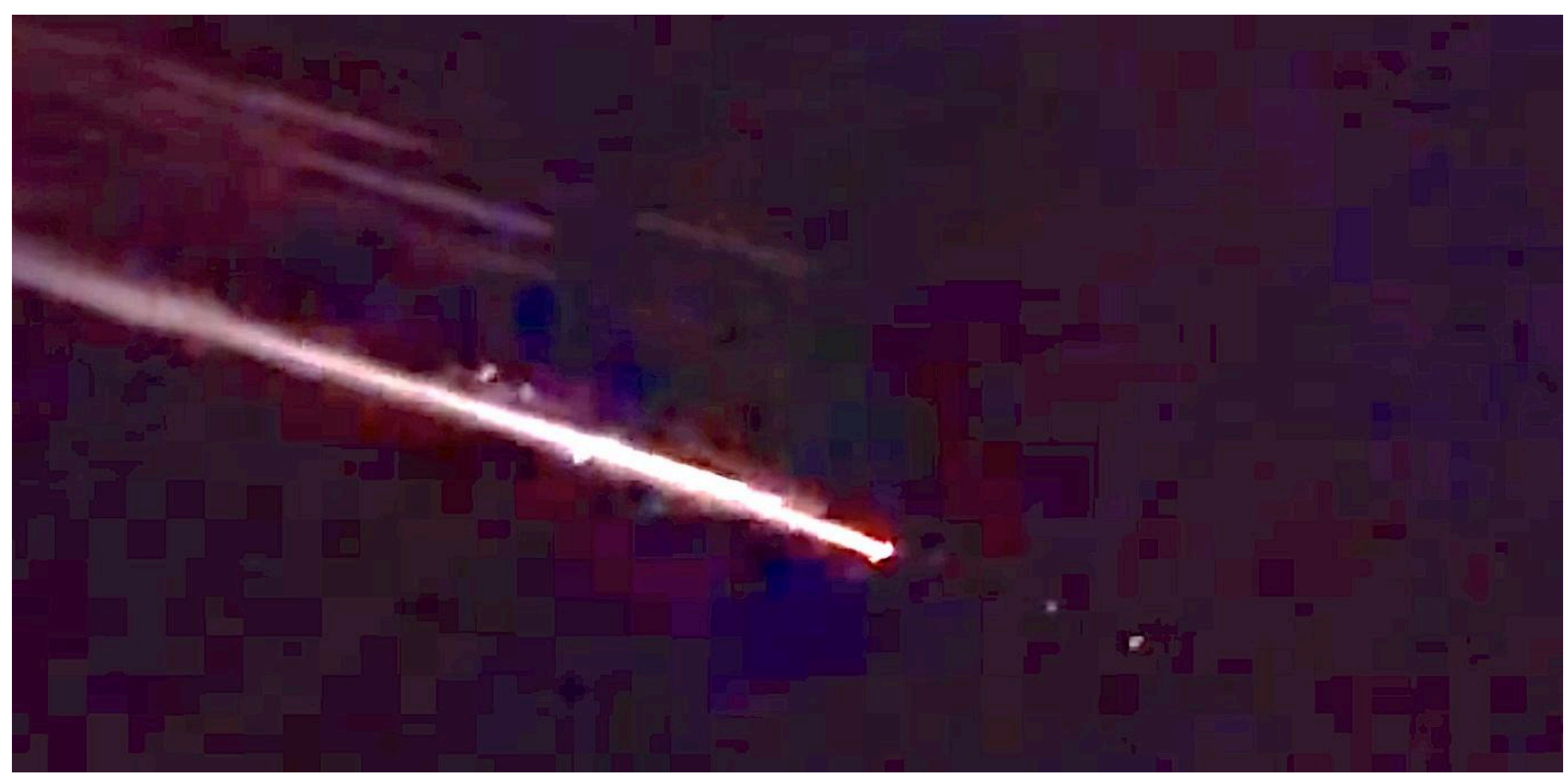

Figure 5. This re-entry of Starlink-5693 on 29 January 03:13 UT over Chicago was photographed by many[4]. With 50 launches and 50 deorbits per day, which will be required to maintain a population of 100,000 satellites, the night sky will be dominated by comparatively bright streaks of uncertain ephemerides. Even with the current 12,000 satellites in LEO, there have been many reports of unusually bright satellites. [33]

## 5. Satellites orbiting above 700 km contribute the most to a dangerous orbital environment

Satellites in high orbit have a high likelihood of colliding, creating a debris runaway cascade effect [17,24,27,28]. High-altitude satellites therefore hasten the existing problem of an increasing night sky background brightness level due to small space debris [20]. In addition, satellites in higher orbits are illuminated for a longer portion of the night (and are thus more likely to appear in images), will leave brighter streaks due to both slower orbital speed and being less out of focus, and will have significantly longer lifetimes before deorbiting in the event of a malfunction.

The estimated increase in space debris 10 cm and larger from satellite constellations is a strong function of altitude and solar activity. A coarse approximation is that collisions with this size of debris can cause a catastrophic collision, while collisions with smaller debris can cause damage but not a catastrophic destruction with generation of large amounts of debris. However, it should be noted that collisions with small debris (< 10 cm) can still debilitate a satellite, producing a non-maneuverable or derelict object, which can itself be a hazard for further catastrophic collisions.

The dependence of space object lifetime on altitude and solar activity is reasonably well understood and is described in [27]. Space object lifetimes increase dramatically as a function of orbital altitude, particularly above about 700 km, increasing to decades and even centuries. As a result, space debris can accumulate much more rapidly at higher altitudes. New research [28] provides estimates for constellations launched into orbits with altitude < 700 km versus those launched into orbits > 700 km. Even if there are considerably fewer satellites in the higher altitude constellations, the production of debris is significantly greater because of the longer orbital decay times. This is due to the fact that derelict satellites in higher orbits (whether from failure of post mission disposal or due to debilitating collision) remain much longer due to the lower atmospheric drag. The results from [28] show a particular problem for constellations at altitude 1000-1200 km. Even lower orbits present demise challenges. There is an *incredibly high* change in expected demise time between operations at 450 km vs. 700 km. Surprisingly, a demise from just 470 km can take *twice as long* as from 450 km in low solar weather conditions.

[4] https://aerospace.org/reentries/55482

Constellations like SpaceX Starlink and Amazon Kuiper, which typically operate below 650 km, should not generate a long-lived debris problem due to the 11-year solar cycle, which tends to clear out the lower orbits. Other constellations such as Guowang, Hughes HVNET, and Eutelsat OneWeb are more problematic in this regard since some or all of their satellites operate (or are proposed to operate) at 800 km or higher.

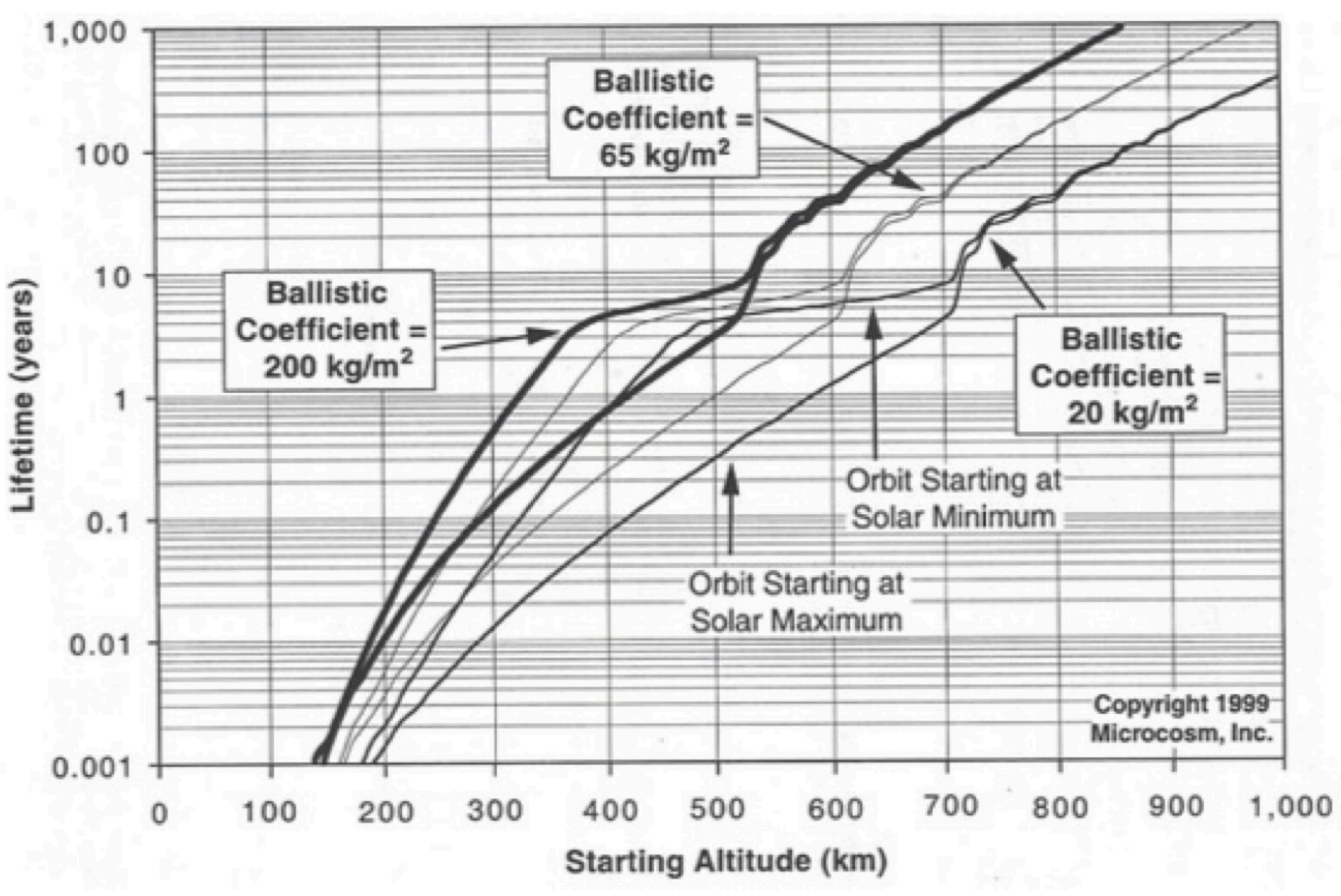

Figure 6. Lifetime of space objects versus initial altitude for maximum and minimum solar activity levels and a variety of ballistic coefficients from [27]. The ballistic coefficient of space debris of 10 cm in size can be in the range 20-65 kg/m$^2$.

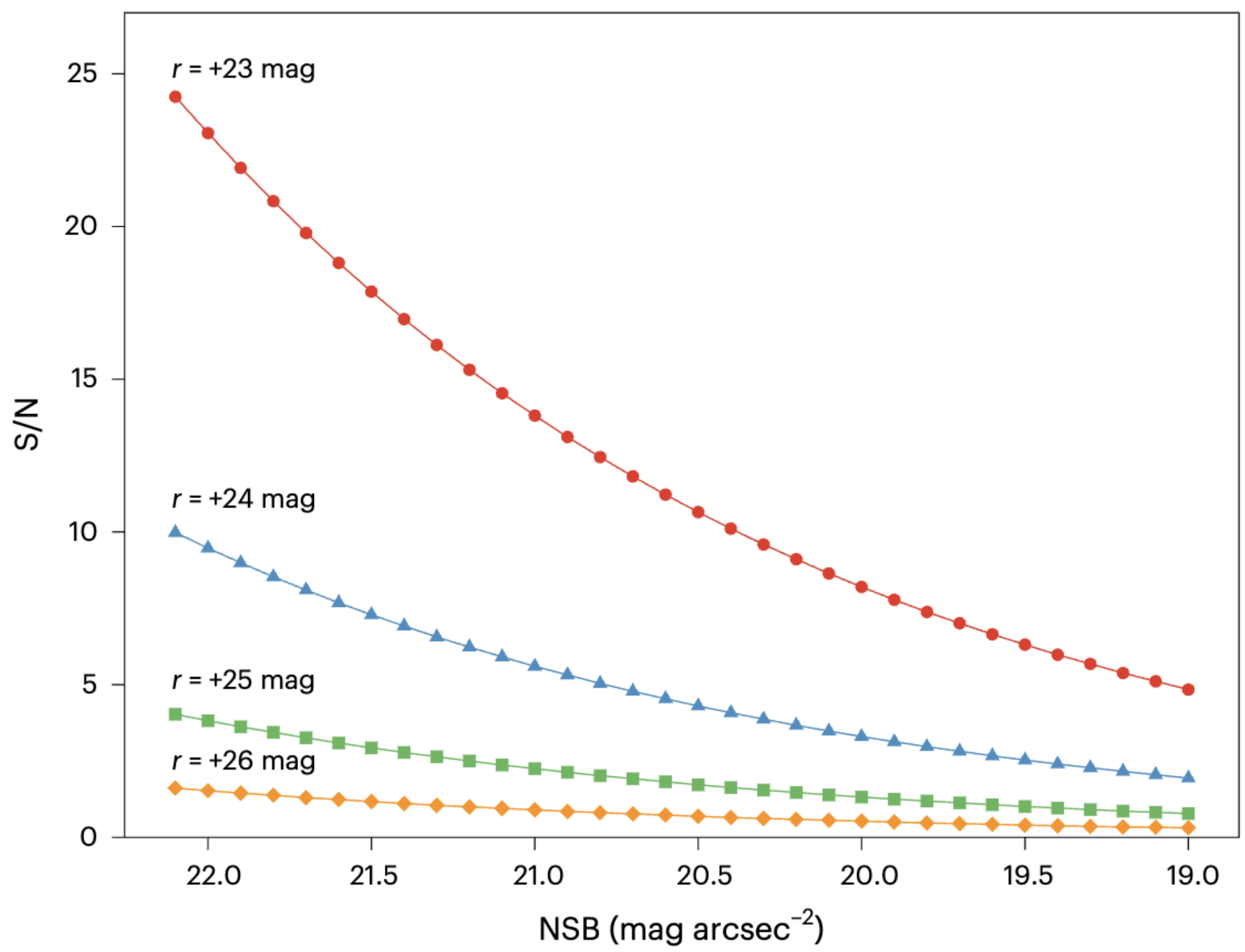

Figure 7. Signal-to-noise ratio (S/N) as a function of background night sky brightness (NSB) for point sources as seen by Rubin LSST, from [20]. Four faint astrophysical sources with slightly different magnitudes near the detection limit are shown, all with a fixed representative exposure time of 20 seconds. As NSB increases (to the right), longer exposure times would be needed to reach any particular desired S/N. The paper notes all log-decades in debris size contribute about the same amount of increased NSB, so debris-generating events are expected to lead to a rapid rise in NSB along with increased serious collision risks.

## 6. Publicly sharing timely satellite ephemerides remains essential

Knowledge of accurate satellite ephemeris data is vital to the suppression of contamination [1,7,8,19]. Including uncertainty information in the form of covariance matrices, and using the recommended standard OEM format for the ephemerides, is strongly recommended. The LSST feature-based scheduler can in principle avoid the ~1000 brightest satellites [35], but this requires accurate timing for both the satellites in question and the planned observation blocks.

A new software tool developed by the IAU CPS SatHub, called SatChecker, enables satellite position prediction[5]. At present, however, it relies on two-line elements (TLEs), which are insufficiently accurate or precise to enable active optical satellite avoidance mitigation techniques, or to definitively identify the source of all past streaks. This is part of an NSF SWIFT-SAT funded project (AST-2332735), and is designed to work with higher precision ephemerides when they are publicly available through TraCSS[6] or a similar API.

[5] https://satchecker.readthedocs.io
[6] https://space.commerce.gov/traffic-coordination-system-for-space-tracss/

Sharing timely and accurate satellite ephemerides is also crucial for coordinating collision avoidance between satellites [32]. The previous section discusses in more detail the risks and impacts of the runaway cascade effect from on-orbit collisions.

## 7. Two distinct regimes, persistent streaks and transient glints, can each create different problems in astronomical images and scientific investigations

Diffuse reflected light from artificial satellites manifests as streaks in astronomical images. Transient specular reflections can look more like point sources. Uncontrolled objects like spinning rocket bodies have more complex phenomenology. The suppression of streaks in co-added images is a different problem than the suppression of glints in single images [4,5,7,16,17]. Glints may appear as a string of dots along a trail, often due to a piece of tumbling debris, or they may appear as one-off transient bogus events, which are more challenging to distinguish from astrophysical transients. For a recent example, see [37].

Glints from satellites and debris in all orbits, even as far out as geosynchronous (of order 35,000 km altitude), may pollute LSST time-domain science. Concave surfaces can produce unusually large reflectivities (BRDF) so that even small satellites at large range can impact LSST. We note that a large population of sub-second glints may severely impact LSST transient science [1,4]. During LSST's typical 30 second exposures, short (sub-second) bright glints are "averaged down" in brightness; in other words, the total sky brightness level continues to rise during the 30 second integration, while the brief glint flux does not. For example, a population of 16 mag 0.5 second glints would appear at 20.5 mag in an LSST exposure, which would be readily detectable as bogus sources. To date we have insufficient LSST transient observations to detect such a population.

## 8. Pixel loss is an insufficient metric for assessing scientific impacts due to non-astrophysical signals from artificial satellites

The impact on LSST science depends not only on the brightness of satellite streaks, but also the number of them. An average LSST exposure will have one or more streaks. The fraction of pixels in the streak and its low surface brightness (LSB) extended wings will be a few percent of the full focal plane. Few science cases depend on covering a specific part of the focal plane. Most science is instead affected by systematic errors introduced by the LSB light.

While recommendations exist that set a target astronomical brightness of fainter than 7th magnitude for any individual satellite, it will be beneficial to also construct metrics that evaluate the overall impact of entire constellations [7,8,9,10,11,21], and to consider how different scientific regimes may be disproportionately harmed by different kinds of contamination. For example, discovering new asteroids is more affected by the large numbers of streaks and glints at low elevation angles, while measuring cosmic shear is more affected by systematic errors that can result from bright linear features when measuring low surface brightness objects.

## 9. Faint objects near the noise detection threshold are particularly susceptible to pernicious systematic errors caused by satellite contamination

The majority of sources detected in any astronomical survey are close to the noise detection threshold, and print-through of uncorrected satellite contamination, even at subtle light levels, can produce systematic errors in survey science (see figures below) [3,12,14].

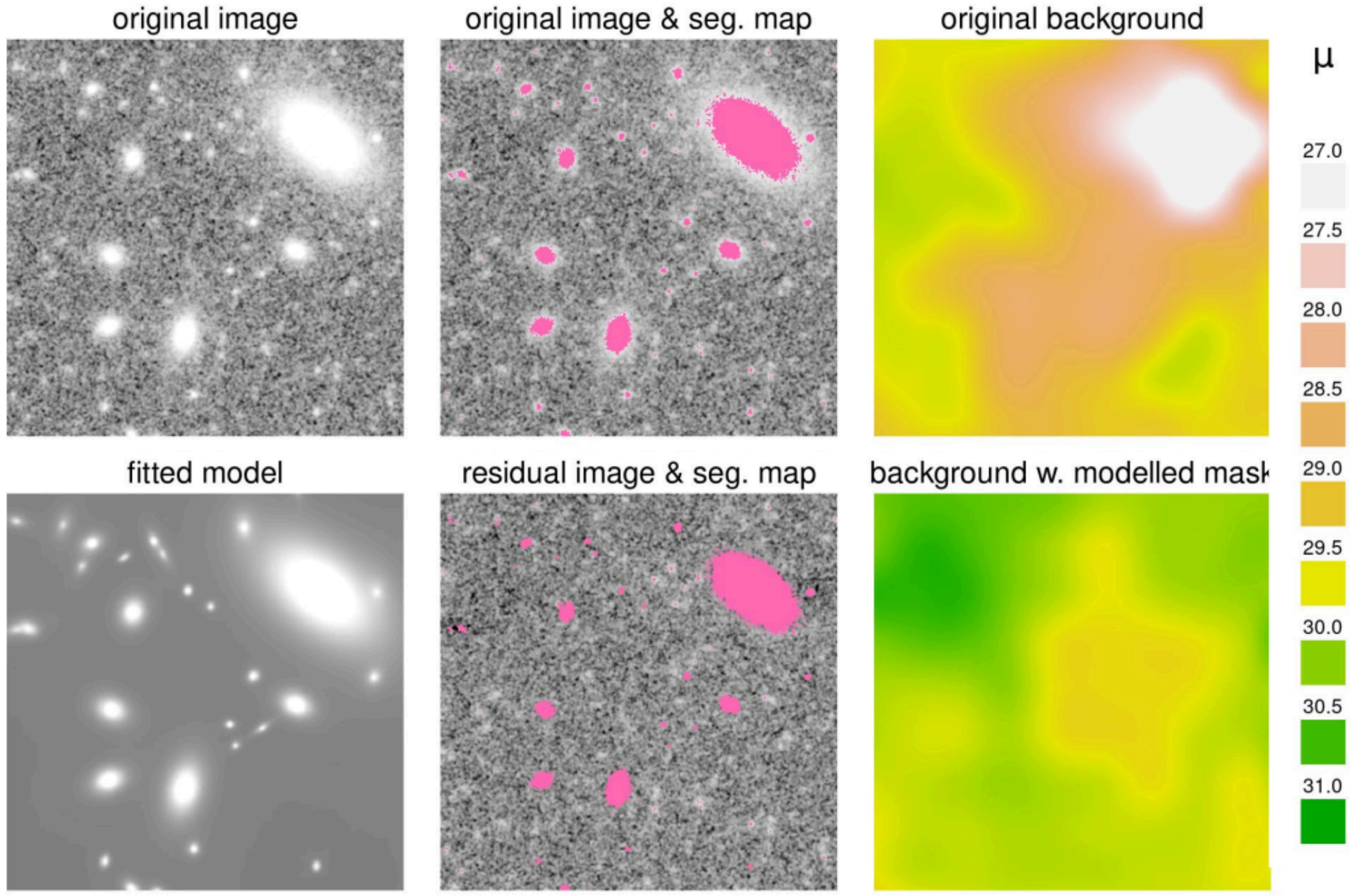

Figure 8. Potential contamination from astrophysical sources can lead to systematic errors if not properly accounted for. Here, an image containing simulated galaxies (top left) is processed using contemporary astronomy tools. Detected sources (*top center*) insufficiently extend to encompass all of the light from the largest of these sources. As a result, the estimated surface brightness backgrounds μ (*top right*) are biased high by several magnitudes, preventing further analysis of the low surface brightness Universe. One such algorithmic approach to mitigate this effect is to generate models which accurately describe large sources of light (*bottom left*). Doing so allows us to more accurately account for light in the outer parts of these sources (*bottom center*) and determine an unbiased background estimate (*bottom right*). Figure adapted from [14].

The faint low surface brightness sky (example figure below) is extremely sensitive to contamination from light sources. While the astronomy community is keen to develop new techniques to mitigate such impacts [14], it is essential that satellite operators continue efforts to minimize contamination of the night sky from satellite streaks. Within the LSST Data Management team, continued efforts should be made to produce a catalog which may be used by the LSB science community to extract information on the positions, shapes and sizes of low surface brightness structures such as dwarf galaxies. This should complement the LSB-focused deep co-added imaging and good seeing co-added imaging already being produced by LSST.

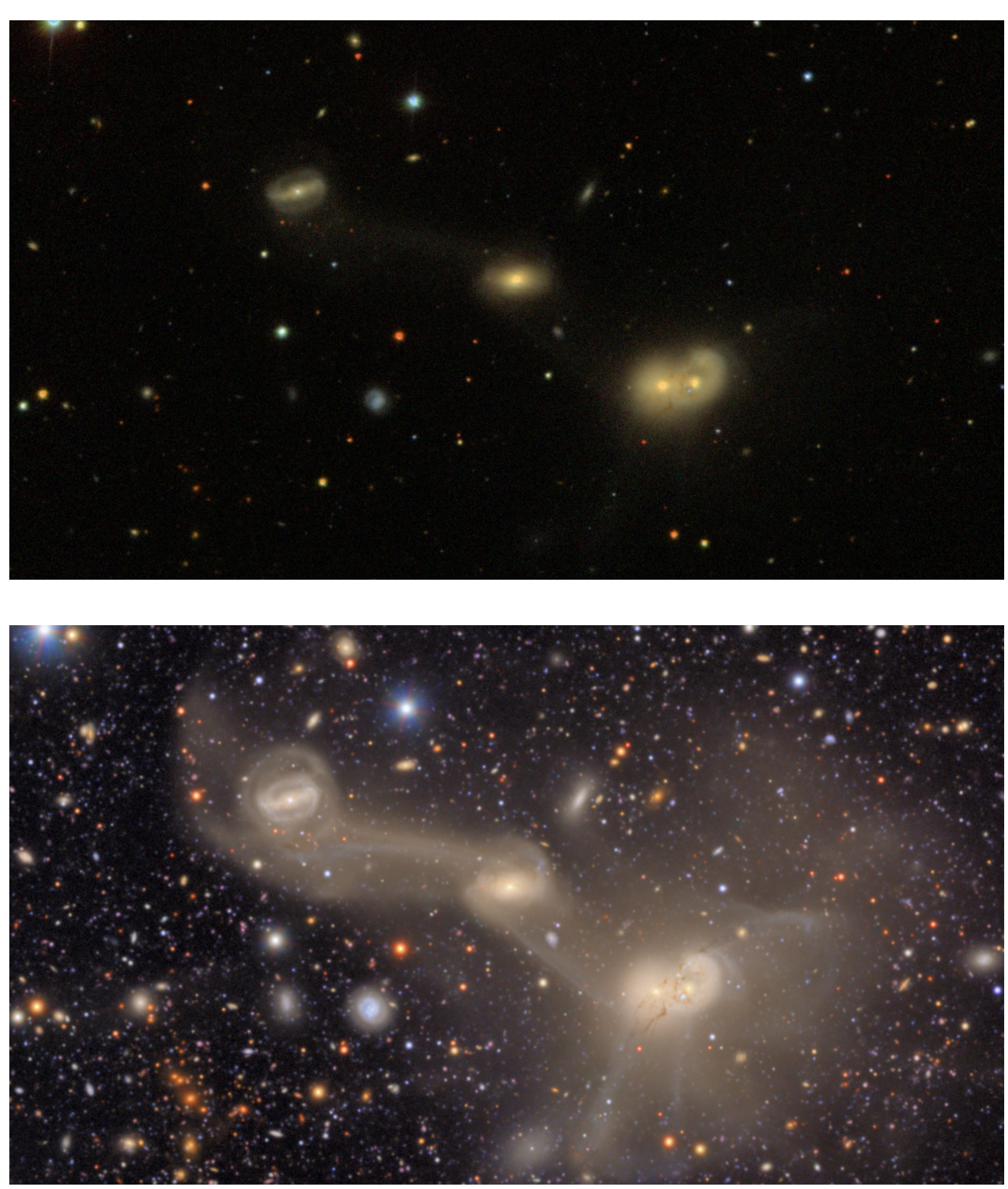

Figure 9. Two images of the interacting galaxy group NGC 4410 in the constellation Virgo. *Top:* An image produced by the Sloan Digital Sky Survey (SDSS), circa 2000. *Bottom:* An image of the same patch of sky produced by the Rubin Observatory, released as part of the Rubin First Look data release, June 2025. Rubin Observatory data shown here is ~100 times fainter than the reference SDSS image above and ~250 times fainter than the night sky in general. Full 10 year depth Rubin imaging will probe yet deeper still. For the first time, Rubin will unmask the low surface brightness Universe on statistically significant volumes of data covering much of the celestial sphere. The potential for scientific discovery is vast, however, the faint Universe is extremely difficult to capture and process and exceedingly susceptible to contamination from spurious sources of light such as satellite streaks.

Systematic error at low surface brightness can, in principle, bias cosmology. In the figure below, faint light spilling over from the edge of a masked satellite streak gets blended with nearby galaxies, creating a coherent line of galaxies that are stretched along the satellite streak direction. This can produce a weak lensing shear bias [12], which may carry through to co-added image catalogs and subsequently to weak lensing cosmology due to imperfect masking [6]. The effect, however, remains within 1-sigma of the confidence limits on the cosmological measurements, and it may be nulled by trimming the catalog of galaxies within two arcsec on each side of 60 arcsec masks.

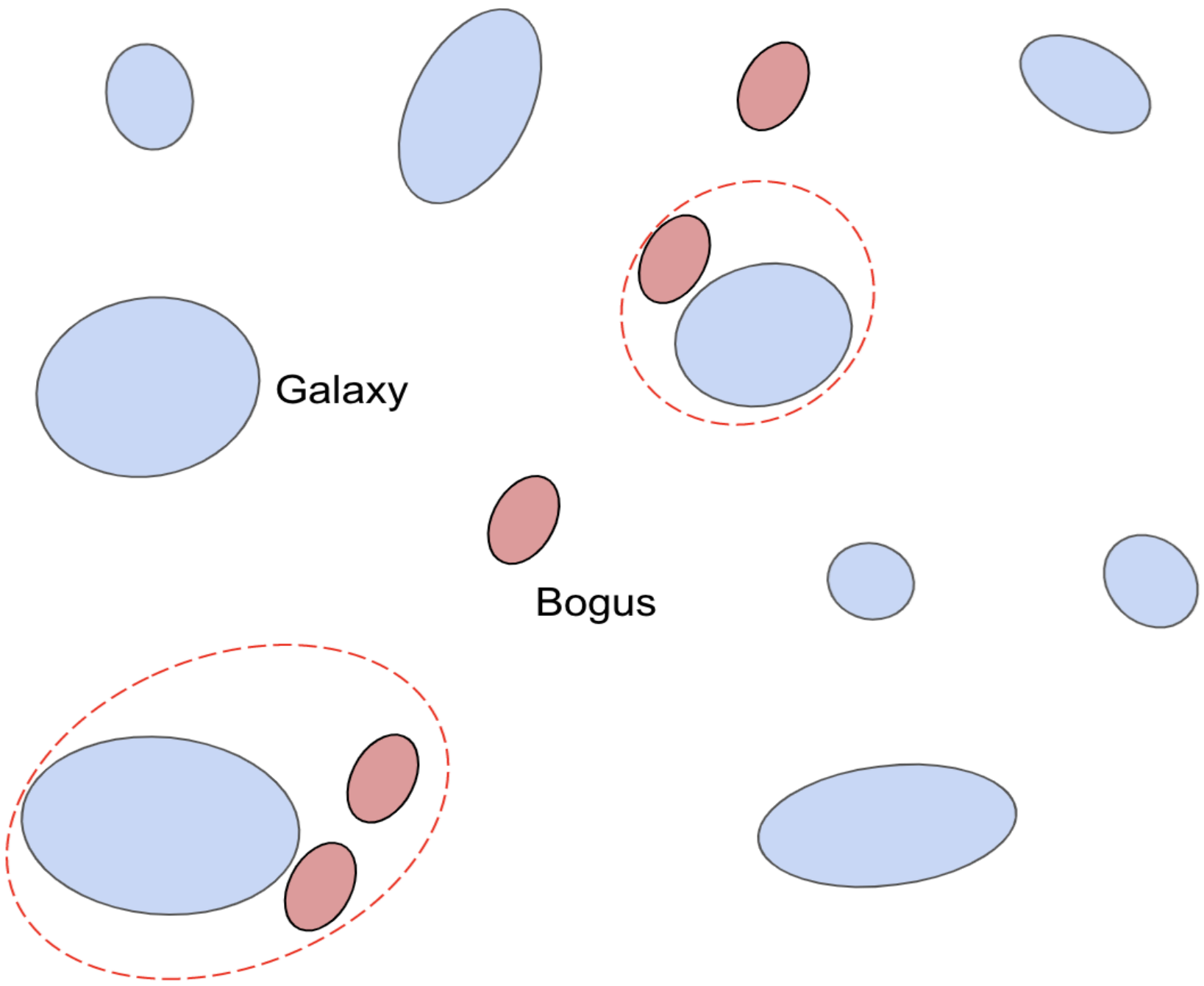

Figure 10. A schematic illustrating how real galaxies (blue) may be contaminated by faint bogus detections (red) due to artificial objects in Earth orbit due to spill-over light at the edge of a wide streak mask. Similar systematic errors occur in single visit image processing, thereby biasing time-domain astronomy. The only mitigation in this scenario is to accurately detect or forecast the expected location of bogus detections and eliminate nearby stars and galaxies from being cataloged at all.

## 10. Continued engagement with policymakers, international organizations, and satellite companies is important to establish and maintain best practices

Certain LEO satellite constellation operators have worked in cooperation with the astronomical community to mitigate reflected sunlight [3,8,9], and the FCC now typically requires coordination with NSF for satellite licenses in the US[7]. Such coordination agreements exist with SpaceX, Amazon, AST SpaceMobile, and PlanetLabs. However, each of these apply only to US-based satellite operations and do not fully mitigate the human and scientific impact of these systems. The EU does have a space policy with clear regulations, and repercussions for failing to protect

[7] e.g., https://www.nsf.gov/mps/updates/nsf-ast-spacemobile-establish-satellite-coordination

the Dark and Quiet Skies are under development[8]. The UN has also begun work toward Dark and Quiet Skies best practice guidance to administrations on policies with a five-year agenda item at the Scientific and Technical Subcommittee of the UN Committee on the Peaceful Uses of Outer Space (COPUOUS) [34].

The ESA has also recently introduced the Zero Debris framework [17], which contains Dark and Quiet Skies Recommendations that align well with this report's findings and recommendations. In summary, these are:

1. Quantify visual brightness of designed satellite before launch
2. Propose and implement design and operational mitigation actions to reduce visual brightness
3. Follow ITU radio regulations
4. Make available data to support mitigation of impacts to astronomy, including but not limited to brightness data, antennae diagrams, orbital profiles, and predicted and real-time orbital elements

Each of these topics is also an area of active interdisciplinary research, which should continue, as effective satellite mitigation cannot be accomplished with a static checklist.

# Recommendations

## 1. Brightness goals

Satellite operators should endeavor to meet, through a combination of satellite design, construction, and operations, the goal of keeping apparent magnitude fainter than 7 V AB magnitude, in accordance with the IAU CPS brightness recommendation [22] and rationale [21].

Specifically, we recommend that satellite builders and operators limit the observed brightness of satellites to below a specific threshold that corresponds to the level at which crosstalk artifacts in LSST Camera images can be corrected to below the noise level. The thresholds for each applicable wavelength band, calculated using a <u>2.0 meter diameter satellite</u> observed at zenith at an orbital height of 1000 km, are presented as radiant intensity values (Watts per steradian integrated over a wavelength band) in the following table:

---

[8] https://defence-industry-space.ec.europa.eu/document/download/0adeee10-af7a-4ac1-aa47-6a5e90cbe288_en?filename=Proposal-for-a-Regulation.pdf

| Wavelength Band [nm] | 405-552 | 553-690 | 690-817 | 818-920 | 920-1010 |
|---|---|---|---|---|---|
| Radiant Intensity Goal [W/sr] | < 29.5 | < 23.3 | < 20.7 | < 21.6 | < 38.5 |

There are two key considerations for scaling these radiant intensities to other satellites and telescopes. First, satellite size. Larger satellites reflect more sunlight (for a given sum BRDF), so the radiant intensity values should be scaled inversely with satellite overall size; larger satellites must meet more stringent radiant intensity limits. However, for larger satellites, the flux density of light is effectively spread over more pixels. The second key consideration is satellite altitude. The radiant intensity values in the table should also be scaled to other satellite sizes according to the distance to the satellite, which can vary as a function of satellite geometry and orbital properties of the satellite constellation using [1,7]

$$\theta_{\text{eff}}^2 = \theta_{\text{atm}}^2 + \frac{D_{\text{satellite}}^2 + D_{\text{mirror}}^2}{d^2}$$

Where one can assume 0.2–0.5 square arcsec (typical across 400–1000 nm) for the $\theta_{atm}$ term, $D_{satellite}$ is the diameter of the satellite in meters, $D_{mirror}$ is the diameter of the telescope aperture (8.5m for LSST), and $d$ is the distance to the satellite (worst case $d$ = orbit height) in meters. $\theta_{eff}$ is the apparent size of the observed satellite.

In general the luminous intensity is directly proportional to the surface area of the reflective surfaces of the satellite and inversely proportional to the square of the distance of the satellite from the observing telescope. Consideration must also be given to the specific properties of the satellite BRDF and the angular geometry at which the satellite is observed, when determining the extent of brightness mitigations needed. The BRDF can be tuned to meet the brightness goals.

See Appendix A for full details relating satellite emission in watts/steradian summed over each wavelength band to the BRDF. Additional examples of satellite size and orbit heights are given.

As guidance for satellite builders, the summed 2-D BRDF of each satellite component can be adjusted to meet the maximum radiant intensity goal shown in the Table above scaled by the satellite size. We note the Table is specifically tailored to the wavelengths that fall inside Rubin Observatory's filter bands. Operators should take care to avoid simply shifting emissions into the infrared or thermal (sub-mm) regime, because other astronomical facilities are sensitive to these wavelengths, and satellites already tend to be thermally bright.

As discussed in [4], this model does not apply to all satellite materials. One notable example is satellites protected by multi-layer insulation (MLI). The as-built MLI geometry is often crinkled, consisting of many 1–10 cm concave surfaces. These can focus sunlight directly onto the observer, producing a bright flash in addition to the ultra high average BRDF.

## 2. Orbital altitudes

We recommend operating constellations at lower altitudes to attenuate impacts on optical astronomical science [3]. A satellite at a lower altitude will have a brighter apparent magnitude than an identical satellite at a higher altitude due to the $1/r^2$ effect, but two other factors nearly cancel out the effect on peak surface brightness detected by observatory cameras:

1. Satellites are more defocused and spread light over more pixels, decreasing electrons/pixel
2. Satellites have a higher angular velocity across the focus plane, so the effective pixel exposure time is lower, decreasing electrons/pixel.

The main benefit of operating constellations at lower altitudes is the decrease in the number of satellites entering the focal plane (~40% fewer from 550 km to 350 km altitude for the LSST focal plane [3]).

We therefore recommend orbits lower than 700 km for all LEO operations, with a preference for the lowest possible orbital altitude. We recognize that lower satellites will leave wider streaks, which increases total pixel loss, but for Rubin Observatory, lowering a streak's peak surface brightness is more important.

## 3. Support coordinated operational and observational experiments

Conducting experiments with varying satellite orientation and internal configuration (such as pointing of solar arrays and antennas), while meeting operational constraints, especially while over-flying observatories, will be helpful in furthering our collective understanding of how to minimize impacts.

During orbit raising, make adjustments to solar panel deployment, angles, and operations to reduce brightness during this phase.

We recommend a suite of experiments be conducted jointly with the astronomical community and the satellite operators. This will require upfront agreements on operational data sharing so results can be published and referred to by other operators and interested parties. These studies would benefit from an array of cameras on the ground, spaced up to a few hundred km apart. (See Appendix B). For example, cooperative experiments between the astronomical community and satellite operators could use ground-based measurements of pirouetting satellites to efficiently map out the reflectance function of individual satellites, aiding in the understanding and reduction of scattered sunlight.

## 4. Coordinate deorbit strategies

Operators should consider, and discuss with the stakeholder communities, a deorbiting strategy that chooses windows in both location and time to minimize deleterious effects. This might include simultaneous deorbiting targeting a single location in the Pacific Ocean, for example. Minimizing reflected sunlight while both raising and lowering orbital height is an important goal.

We recommend the highest luminosity burn-up events be timed during the day over the Pacific Ocean. It is also important to ensure objects do fully burn up and do not impact the ground. We further encourage operators to minimize the number of inoperable “post-mission” satellites by allowing no more than 20% of an operator’s total orbiting satellites to be in post-mission status.

## 5. Share timely ephemeris and attitude information in publicly accessible locations following international standard and precedents

We recommend imposing a requirement that all constellation operators make accurate and up-to-date ephemeris information publicly accessible. This is important not only for astronomical image processing, but also for space traffic management [32].

If operators publish (bulk) BRDFs of their satellites, or BRDFs of their major reflecting surfaces, predictions of when their satellites will be brightest could be included in LSST observing schedule considerations. For satellite avoidance with LSST, a maximum error in the predicted position of 1 second and 1 arcminute on timescales of 30 seconds as well as an uncertainty in the predicted brightness of 0.5 magnitude would be effective. We note this is sufficient only to avoid pointing at the brightest satellites. To effectively and proactively identify the position of a satellite trail or glint within a single LSST camera detector image, on-board GPS-level precision is required, particularly in the along-track direction in both space and time.

We recommend that satellite operators share satellite attitude information publicly as a function of time, so that astronomers can understand when it will be bright. We also recommend sharing precise ephemerides in a publicly accessible way: use the orbital ephemeris message format (OEM) and include uncertainties (covariances).

## 6. Develop new constellation-wide metrics and standards to assess mitigations

We recommend further efforts to establish constellation-scale metrics (glint rates, aggregate streak statistics, etc.) to allow a comparison of the astronomical consequences of various orbital and operational options. One promising avenue involves injecting synthetic streaks into real LSST images and studying the differences in source detection in the LSST Science Pipelines.

These constellation-scale metrics should be in addition to, not in place of, per-satellite limits (e.g., brightness). Such metrics can quantify important cumulative factors that are impossible to assess on an individual satellite basis. For example:

- A constellation with fewer larger satellites is likely to have a similar overall impact on astronomy as a constellation with more smaller satellites.
- A constellation with satellites at a wide range of orbital altitudes will have a different impact than if the same constellation were operated in a single orbital shell.
- A constellation that has a high glint rate near twilight but very few streaks visible near midnight will have a different impact than if the same number of glints were evenly distributed in time.

We encourage continued work by the IAU CPS and other groups to bring together experts representing multiple countries, satellite operators, and astronomical institutions to collaborate to develop such metrics that may be applied to satellite constellations of various sizes. Some starting points may be found in [22,34].

In addition, we strongly endorse ongoing work by the IAU CPS Policy Hub, SatHub, and Industry & Technology Hub toward developing recommended standards for optical brightness characterization (i.e., how satellite operators should measure brightness), including aggregate impacts. These standards should include (1) best practices in data reduction, (2) requirements to meet or exceed a minimum set of reported information of observational results, and (3) test/fail metrics for determining the validity of lab experiments and simulations when compared with measurements taken when satellites are in orbit. They must be more broadly applicable than the scope of this workshop, i.e., taking into account the needs of other observatories and LEOsat companies. Finally, they should serve as a complementary set of guidelines alongside mitigation targets (e.g., brightness limits) that seek to foster cooperation, trust, and innovation.

## 7. Super bright satellites should be discouraged, because they are a problem even if there are fewer of them

There are some proposed satellites which particularly threaten nighttime optical astronomy due to being inherently extremely bright. One example is large reflective surfaces designed to collect solar power. This has its origin in 1978[9] (and even earlier, in 1929, as an idea[10]). The following is a quote from a panel report on the 1978 proposal.

> "Power satellites in space would be expected to reflect substantial amounts of light. Even with a coefficient of reflectivity as low as four percent, each power satellite would appear to be brighter than all but the brightest bodies in the sky (the sun and the moon) and would be about as bright as Venus when it is most visible. Multiple satellites would brighten the sky considerably. For example, 60 satellites would provide as much light as the moon between its new and quarter phases across a band 40° long and 10° wide. Earth-based optical observations would be hindered under these conditions."

Over the years there have been many other proposals, from space advertising to artificial lighting from space. A recent example is Reflect Orbital. They plan 4000 satellites by 2030, each 18x18 meters which will illuminate selected ground locations several hours past twilight, as well as illuminate selected areas all night. This is potentially ruinous to ground-based astronomy. We recommend working together with any such proposers to characterize and minimize the reflected sunlight onto the ground outside of specifically targeted regions during all phases of operations.

## 8. Detect and flag the presence of streaks and glints in all affected LSST data products

The Rubin Observatory project should fully investigate the impact of complex streaks and glints, and how they propagate into catalogs, difference images, alerts, and co-added images. This should include robust techniques to distinguish satellites from natural Solar System objects, cosmic rays, and other astrophysical phenomena.

[9] https://ntrs.nasa.gov/api/citations/19820014802/downloads/19820014802.pdf
[10] https://en.wikipedia.org/wiki/Sun_gun

We note LSST Alerts will not be issued for known satellites in compliance with [26]. To accomplish this, satellite TLEs provided by Space-Track.org will be used to exclude any Rubin alerts where the source bounding box contains pixels which overlap with known satellites. Additional streak and glint trail detection, as well as long trailed source filtering, will be used to exclude additional artificial signals that are likely due to satellites or debris. However, despite these best efforts, it is still likely that signals from artificial objects in Earth orbit will still appear in some Rubin data products.

Therefore, we recommend a combination of Rubin and community effort continue to work toward identifying and characterizing false positive detections and pernicious systematic biases due to satellites and debris. It is important to document which steps are taken for which data products so that downstream users, including Alert Brokers, understand what kinds of satellite and debris contamination are possible in images and catalogs.

If the population of extremely bright satellites grows significantly during Rubin Operations, which is unfortunately likely, the LSST feature-based scheduler should be prepared to avoid regions of sky accurately forecast to contain a list of known bright satellites [35].

While the LSST Data Management team has developed successful streak bit mask detection algorithms for brighter streak trails, it does not perform well for faint streaks or streaks that vary in brightness, and it sometimes detects other linear features (e.g., diffraction spikes). We recommend developing an algorithm that is computationally efficient at detecting streaks and also delivers high reliability in the completeness-accuracy relation. One possible way to achieve this may be a neural posterior estimation process [31]. Recursive training with a wide variety of LSST camera data and sources of truth (e.g., Zooniverse campaign and/or synthetic source injection) are needed to investigate which kinds of streak and glint morphologies are effectively detected and quantify completeness.

## 9. Build a comprehensive database of satellite signatures in LSST data

We recommend the construction of a calibrated comprehensive database of LSST streaks to ultra faint levels. Most LSST discoveries will occur at the faintest levels of surface brightness. This is where LSST is unique: the discovery of the unexpected. Classes of bright objects have been detected in previous sky surveys: known knowns. LSST will be unique in discovering the unknowns.

Satellite streaks can cause systematic errors in both cosmological and time-domain measurements. These systematics arise from bogus detections of “objects” due to incompletely detected faint streaks as well as light spilling beyond masked bright streaks.[6]  LSST dark matter and dark energy science will be directly affected by this, as well as validity of transient object alerts.

Both static deep sky and time-domain data will still contain induced systematic errors. We recommend building a novel automated faint streak detection and analysis pipeline, and a comprehensive database for catalogs and images.

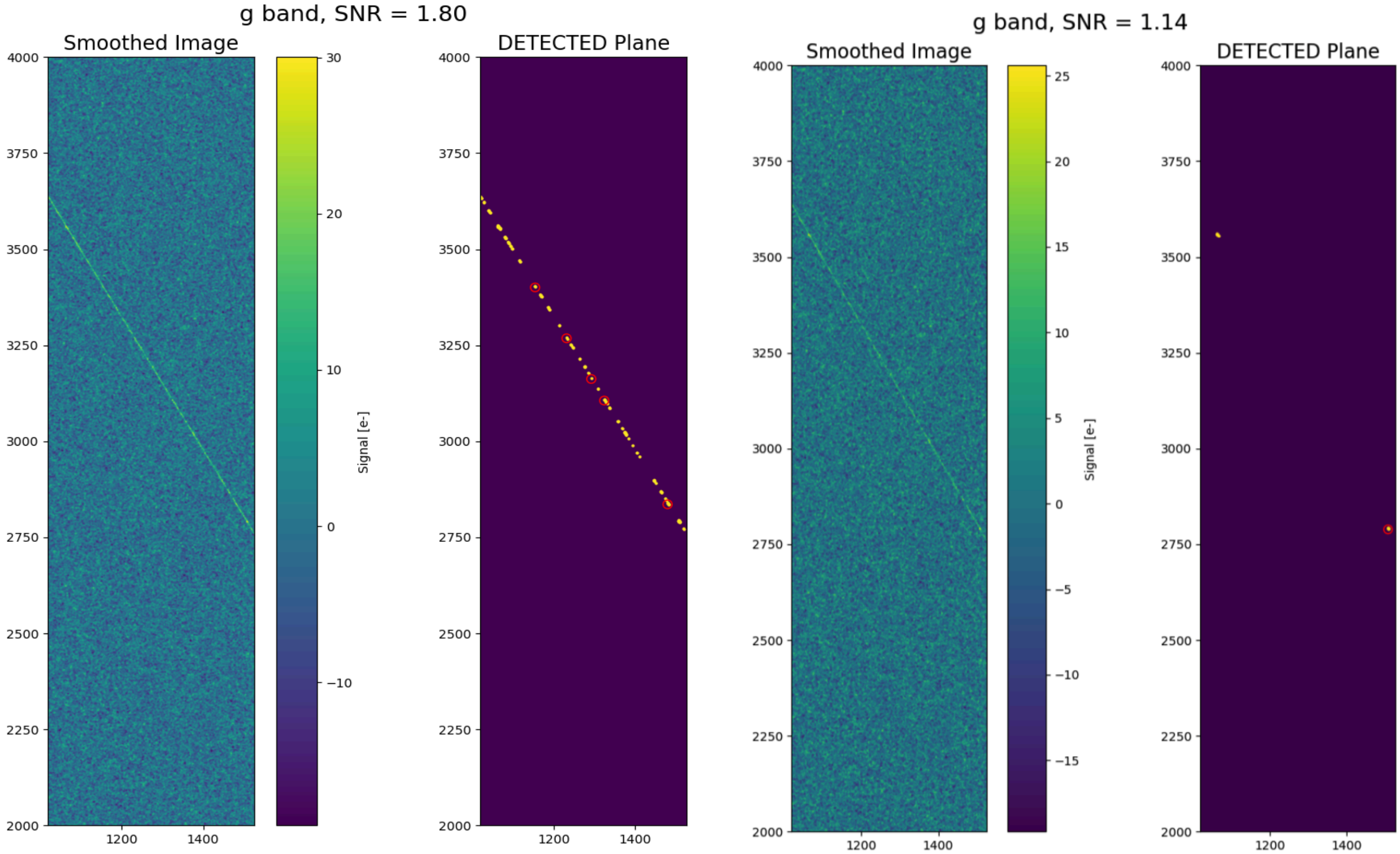

Figure 11. Examples of typical low signal-to-noise streaks as seen in LSST CCDs. In one of the 16 segments of an LSST CCD, a S/N = 1.8 streak is incompletely detected, resulting in a group of isolated bogus catalogued objects (*left panels*). As the S/N decreases to ~1 near the surface brightness detection threshold (*right panels*), only one or two individual sources remain, resulting in two bogus detections. Because they are no longer obviously part of a linear feature, typical streak or glint detection algorithms would not flag these sources. This results in bogus objects and catalog incompleteness at low surface brightness.

The astronomical community would benefit from a comprehensive and calibrated assay of detected streaks in LSST data as a function of satellite brightness. This would enable anyone to test their science objective against reality.

As part of this effort, we recommend that the LSST Data Management team continue to develop algorithms which preserve the ultra-faint low surface brightness sky, and commit to providing such information as a data product to be served to the wider scientific community. The LSST has already invested significant effort in developing full focal plane sky correction algorithms which preserve the ultra faint sky on large scales [29,30]. Further such advancements are actively being pursued which correct for any minute sky oversubtraction of the background around extended astronomical sources [14]. The provision of these data to the wider community prior to satellite streak detection and correction will allow for comprehensive studies of satellite impact and further subsequent corrective actions as deemed necessary.

## 10. Formally engage in national and international forums and with community science groups

We recommend continued formal mechanisms for US inter-agency participation in ongoing international forums regarding protocols for large-constellation satellite operations and debris mitigation with active leadership from NSF.

To ensure LSST is still able to detect unexpected events or objects (“unknown unknowns”), NSF with LSST Operations should encourage increased collaboration among all of the Alert Brokers, Science Collaborations, and the LSST Project. It is essential to understand what kinds of signals are expected in LSST data products both from known astrophysical phenomena and from contamination due to satellites and debris.

There currently exists no calibrated collection of ultra faint satellite data, particularly for LSST. LSST Science Pipelines software will be used to detect both bright and faint LEO satellite streaks. These data will be calibrated by random injection of streaks into images. Recursive fractions of those will be used in initial training.

# Appendix A. Brightness Goals

This Appendix expands on Recommendation #1 with more examples of maximum watts/steradian as well as general information for scaling to arbitrary satellite geometries and orbital heights. Some background on how Rubin calculates the acceptable maximum brightness may be helpful [7]. The origin of this radiant intensity limit is that the LSST camera exhibits non-linear crosstalk between the 16 subsections of each CCD. Much of this originates in the electronics following the CCD. While the CCD and electronics have a large dynamic range, the low level response must be emphasized because most of the LSST science is very low surface brightness (a few electrons per pixel). In order to correct for the crosstalk, millions of on-sky images of non-saturated stars must be obtained. The feasibility of that crosstalk calibration sets the limit of correctable crosstalk as shown in the figure below. LSST also has a collimated beam projector in the dome which places a large spot on each of the 16 channels of each of the 189 CCDs. Using a swept laser, approximate crosstalk matrices can be obtained and changes monitored over time.

The maximum outgoing radiant intensity will depend on the properties of the BRDF, however a simple assumption for a “worst case scenario” is that the satellite will appear brightest when observed at zenith, for which the range (distance from the satellite to the observer) is equal to the orbital height and the satellite is closest to the telescope. A satellite that adheres to the brightness goals when observed at zenith is then likely to be dimmer at all other possible observing geometries. We derive brightness goals under this assumption, but it is important to remember that when we refer to orbital height this is in actuality the range, unless specifically noted.

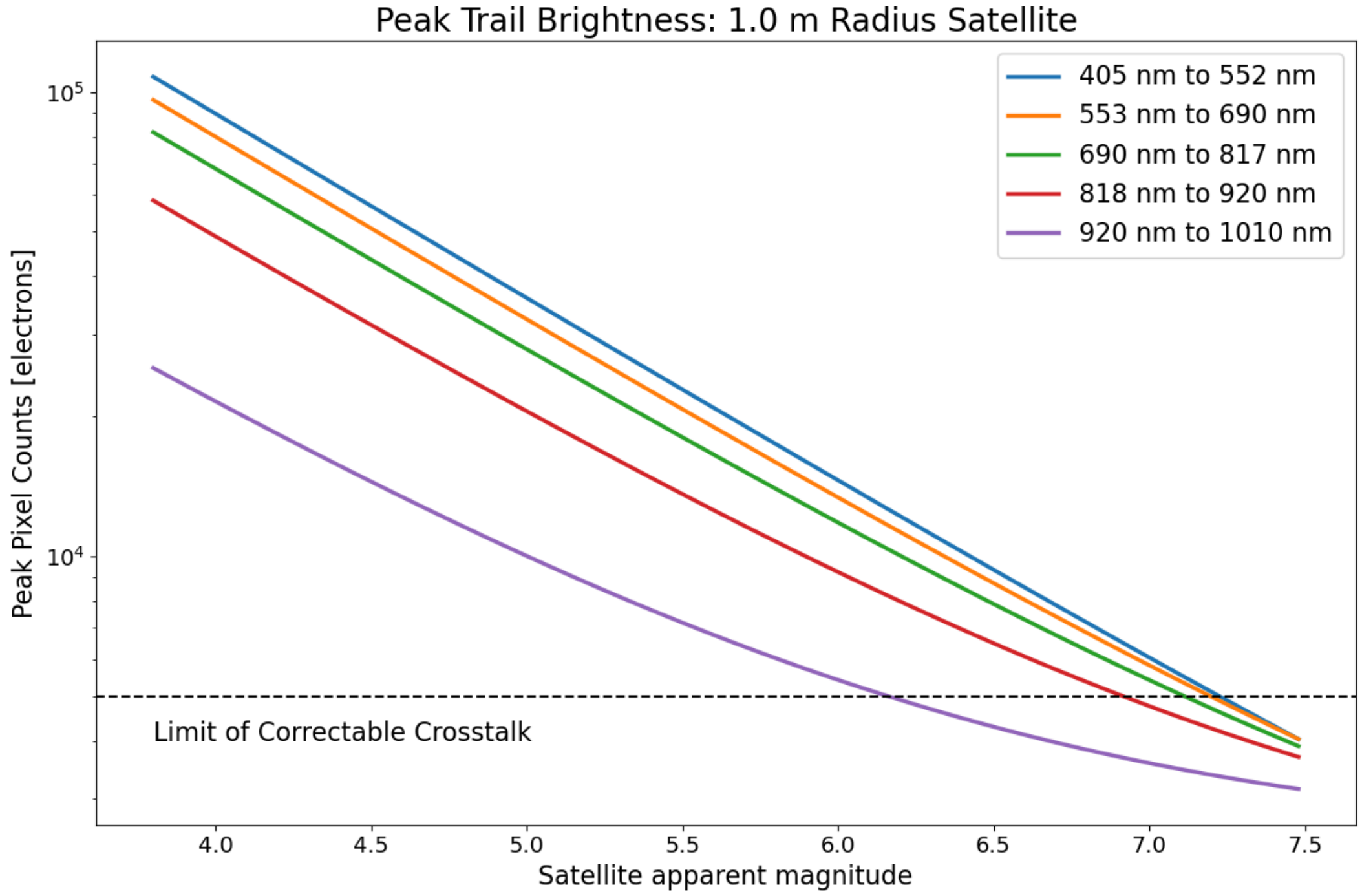

Figure 12. The peak pixel count in each LSSTCam wavelength band (except u-band) for a 2.0 meter diameter satellite observed at a range of 1000 km, as a function of the satellite apparent magnitude, i.e. the brightness of the satellite if it were tracked by the telescope. Also shown is the pixel count limit for crosstalk effects to be corrected below the noise level. Note that because of increased sky emission at long wavelengths, the range of tracked magnitude for 10% precision correctable crosstalk is 6.5 – 7 AB mag.

Using simulation tools, the peak pixel count of a satellite trail can be calculated by modeling the trail profile. The peak pixel counts are first converted to electrons by considering the system gain in electrons per count. Each electron collected by the detector corresponds to a photon; by assuming an effective wavelength for the filter band, the incoming radiant flux can be calculated. Next we calculate the radiant intensity from the radiant flux, which takes into consideration the satellite orbital height and telescope size.

Finally, the total incoming radiant intensity is calculated by dividing by the filter band throughput integral, which includes losses due to atmospheric absorption, the telescope optics, and the detector quantum efficiency.

The result is a goal value that is nearly independent of the observing telescope, but is derived under assumptions specific to the LSST Camera. The table below shows the radiant intensity and radiance goal for a <u>2 meter diameter satellite</u> at two different orbital heights (1000 km and 350 km) that is observed at zenith by the LSST Camera.

2 meter Satellite at 1000 km Orbital Height

| Wavelength Band [nm] | 405-552 | 553-690 | 690-817 | 818-920 | 920-1010 |
| --- | --- | --- | --- | --- | --- |
| Radiant Intensity Goal [W/sr] | < 29.5 | < 23.3 | < 20.7 | < 21.6 | < 38.5 |
| Radiance [W/(sr m^2)] | < 9.4 | < 7.4 | < 6.6 | < 6.9 | < 12.3 |

2 meter Satellite at 350 km Orbital Height

| Wavelength Band [nm] | 405-552 | 553-690 | 690-817 | 818-920 | 920-1010 |
| --- | --- | --- | --- | --- | --- |
| Radiant Intensity Goal [W/sr] | < 24.3 | < 19.2 | < 17.1 | < 17.8 | < 31.8 |
| Radiance [W/(sr m^2)] | < 7.7 | < 6.1 | < 5.4 | < 5.7 | < 10.1 |

These radiant intensity goals are valid only for a specific telescope geometry and satellite orbital height; it is therefore necessary to provide relations by which to scale these limits to other satellite constellation properties.

The geometry of the reflected light from the satellite is shown in the following figure from [1]. Here, a fraction of the incoming illumination from a point source (the Sun) is reflected in an outgoing direction towards an observer (the telescope).

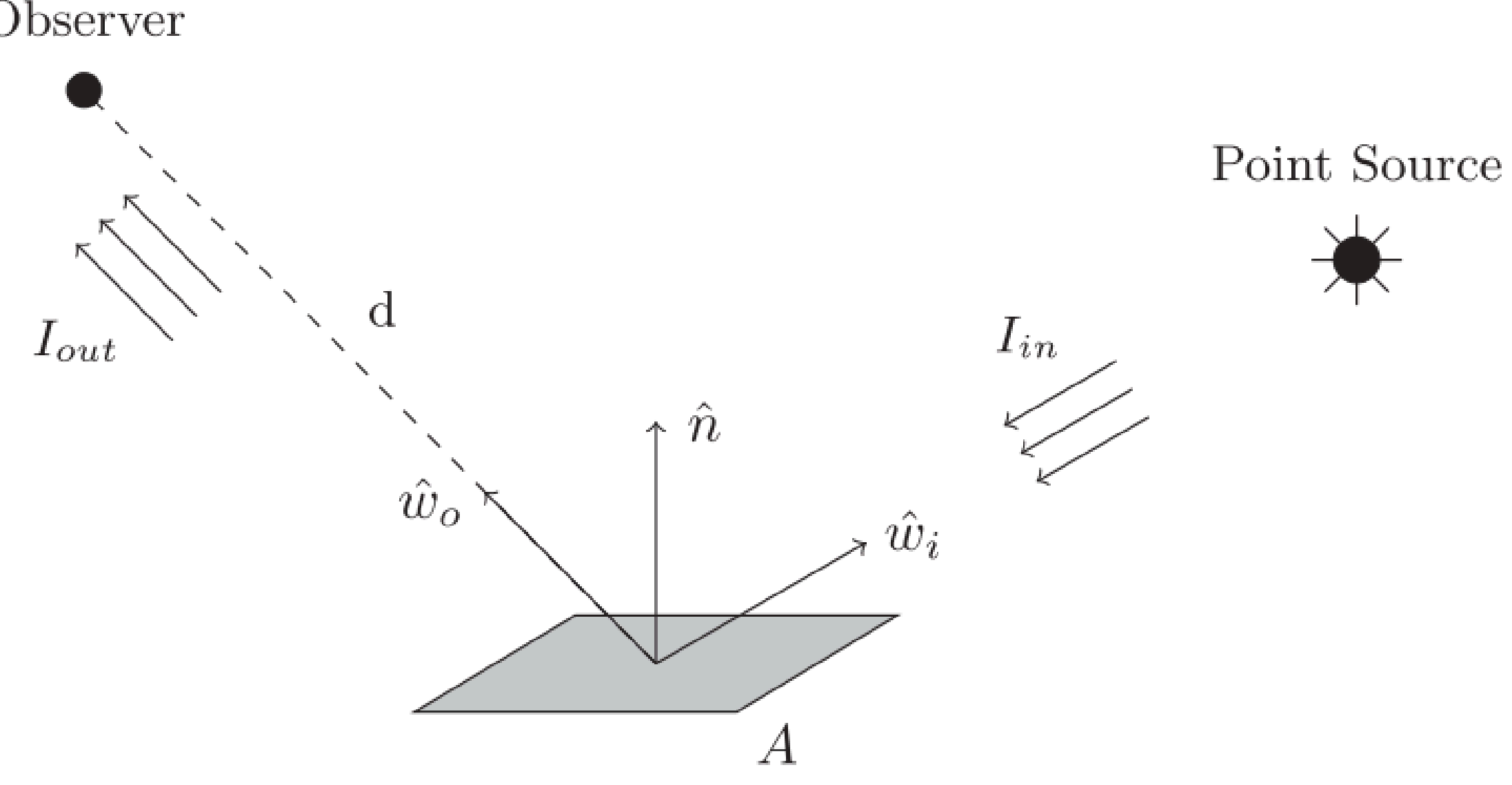

Figure 13. The ratio of the outgoing radiant intensity to the incoming irradiance is a function of the angular geometry, satellite area A, distance to the observer, and the bidirectional reflectance distribution function BRDF.

The brightness of an arbitrary satellite is determined by this fraction and can be calculated using the following equation, outlined in [1].

$$\frac{I_{\text{out}}}{I_{\text{in}}} = (\hat{w}_i \cdot \hat{n})(\hat{w}_o \cdot \hat{n}) f_r(\hat{w}_i, \hat{w}_o) \frac{A}{d^2}$$

This equation can be used to scale the radiant intensity goals to account for changes in satellite geometry, orbital height, and BRDF.

We first consider changes to the satellite reflective surface area. In accordance with the preceding equation, the outgoing radiant intensity will increase linearly with surface area. In the absence of changes to the satellite orbital height, the defocus kernel that is set by the angular size of the telescope pupil will be unchanged. Therefore the effect of changes to the surface brightness profile of the satellite are reduced and are subdominant to the increases in surface area. We therefore recommend that satellite operators simply follow the linear dependence when determining optical emission limits.

As an example for a larger satellite, the brightness limits can be recalculated assuming a 4 meter diameter satellite at two different orbital heights of 1000 km and 350 km, shown in the table below.

4 meter Satellite at 1000 km Orbital Height

| Wavelength Band [nm] | 405-552 | 553-690 | 690-817 | 818-920 | 920-1010 |
|---|---|---|---|---|---|
| Radiant Intensity Goal [W/sr] | < 29.9 | < 23.6 | < 21.0 | < 21.9 | < 39.1 |
| Radiance Goal [W/(sr m^2)] | < 2.4 | < 1.9 | < 1.7 | < 1.7 | < 3.1 |

4 meter Satellite at 350 km Orbital Height

| Wavelength Band [nm] | 405-552 | 553-690 | 690-817 | 818-920 | 920-1010 |
|---|---|---|---|---|---|
| Radiant Intensity Goal [W/sr] | < 27.7 | < 21.9 | < 19.4 | < 20.3 | < 36.2 |
| Radiance Goal [W/(sr m^2)] | < 2.2 | < 1.7 | < 1.5 | < 1.6 | < 2.9 |

For this larger satellite the radiant intensity goals have increased slightly; this is a result of the surface brightness profile on the focal plane becoming slightly wider. However the radiance goals have decreased significantly to account for the larger reflective surface area. This demonstrates the need to decrease the radiant flux reflected per unit area of the satellite reflective surface to maintain a total radiant intensity below the prescribed goal.

Our final example is for a 10 meter diameter satellite. This even larger satellite continues to demonstrate the trend of increasing radiant intensity goals, but substantially decreased radiance goals.

10 meter Satellite at 1000 km Orbital Height

| Wavelength Band [nm] | 405-552 | 553-690 | 690-817 | 818-920 | 920-1010 |
|---|---|---|---|---|---|
| Radiant Intensity Goal [W/sr] | < 32.8 | < 25.9 | < 23.0 | < 24.0 | < 42.9 |
| Radiance Goal [W/(sr m^2)] | < 0.4 | < 0.3 | < 0.3 | < 0.3 | < 0.6 |

10 meter Satellite at 350 km Orbital Height

| Wavelength Band [nm] | 405-552 | 553-690 | 690-817 | 818-920 | 920-1010 |
|---|---|---|---|---|---|
| Radiant Intensity Goal [W/sr] | < 31.2 | < 24.7 | < 21.9 | < 22.8 | < 40.8 |
| Radiance Goal [W/(sr m^2)] | < 0.4 | < 0.3 | < 0.3 | < 0.3 | < 0.5 |

Orbital height has an effect on the surface brightness of the satellite streak in the camera focal plane. For low orbits the effective exposure time per pixel is shorter because the angular velocity of the satellite is higher.[2] When combined with the defocus effect, lower orbits are best.[3] While the orbital height is set by the satellite operator, the range is determined by the satellite constellation orbital conditions at the time of the telescope observation and the angular geometry at which the satellite is observed, both of which can be considered unknown a priori.

Although for the preceding examples we assumed that the satellite was observed at zenith and therefore the orbital height was equal to the range, for completeness we provide a calculation for the satellite range for arbitrary observed angle from zenith. The range can be calculated as a function of the angle from telescope zenith at which the satellite is observed and the corresponding angle from satellite nadir to the observing telescope, which does depend on the orbital height.

$$r = \begin{cases} h & \theta_N = 0 \\ R_{\oplus} \frac{\sin(\theta_Z - \theta_N)}{\sin \theta_N} & \theta_N \neq 0 \end{cases}.$$

We note that convex surfaces also help reduce the brightness. Concave reflecting surfaces should be avoided in satellite design.

Finally, satellite operators and builders can have 2-D BRDF measurements made on components of satellites at a number of companies in the US, Canada, and Europe. This is useful in the design and development phase.  On-orbit photometry of as-built satellites from Earth or space can serve as validation when combined with conops data on satellite and panel orientation.

# Appendix B. Simultaneous observations of reflected sunlight

Assessing the impact of satellite constellations on astronomy will be an enduring challenge. Modeling the footprint on the ground of reflected sunlight is a way to map out the BRDF of a satellite at fixed solar illumination angle, and a fixed orientation and configuration of the satellite. Apart from the overall orientation of the satellite bus relative to nadir and solar illumination angle, the solar panels and any steerable communications antenna are independent degrees of freedom that change the distribution of reflected sunlight. This argues in favor of making simultaneous measurements of reflected sunlight from multiple ground stations. A satellite at an altitude of 350 km is visible from the ground at elevation > 20 degrees within a circle of radius 820 km. One of these satellites that passes directly overhead is visible for four minutes. If we want significant temporal overlap in imaging from multiple sites, a conceptual design would entail installing multiple wide-field cameras at sites within a circle ~<300 km in radius. We note that there are existing Chilean observatories with power and internet that might be appropriate for this. Figure BB shows the locations of (ordered S to N) observatories that are well-positioned to host coordinated cameras: El Sauce, Cerro Pachon, CTIO, and La Silla.

While there are existing all-sky cameras at many of these sites, we note that CMOS technology allows rapid imaging in which streak lengths can be minimized.

Operating this camera array in a fashion that is coordinated with satellite pirouettes would provide credible, far-field measurements of satellite BRDF.

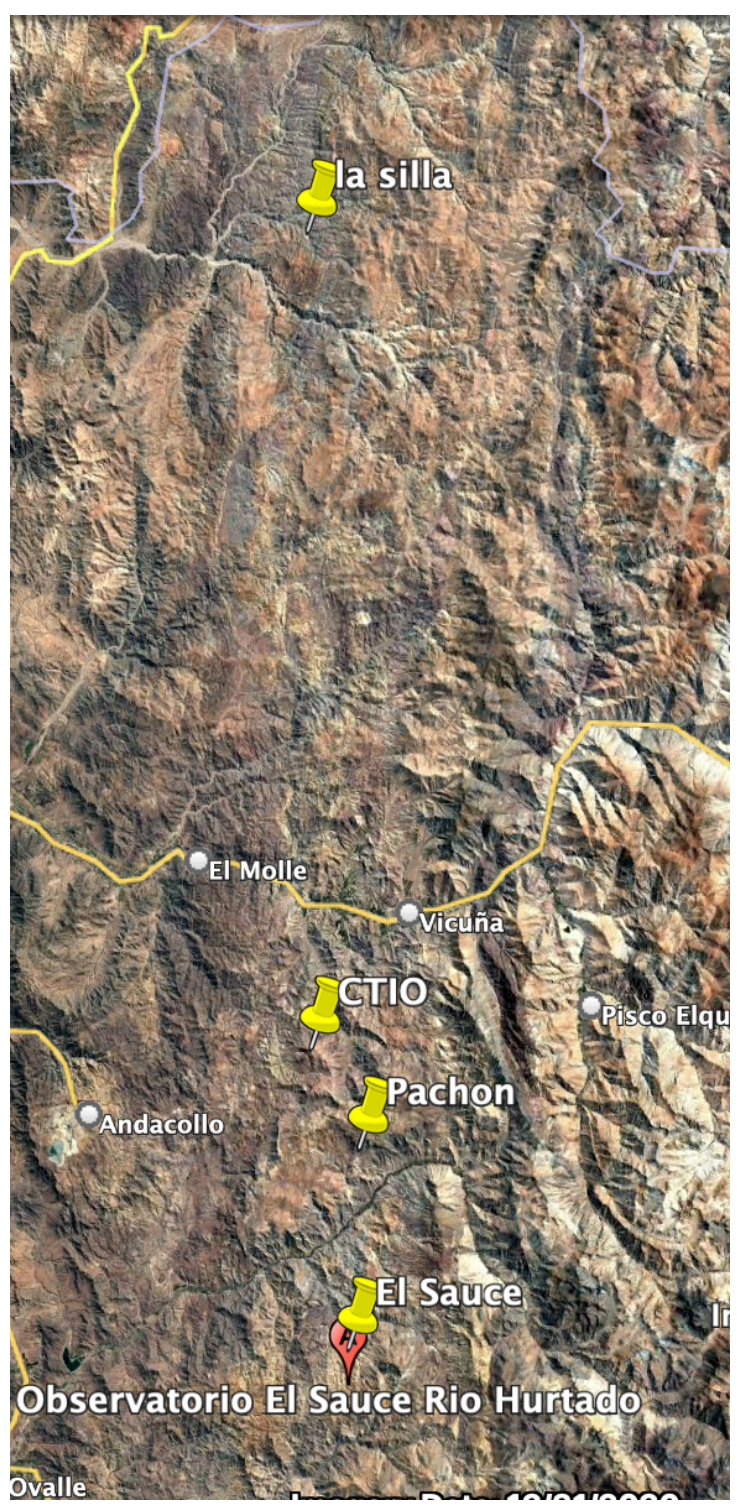

Figure 14. This diagram shows the locations of existing observatories that could host cameras designed to simultaneously image the reflected-sunlight footprint of satellites, on the ground. This will add valuable information to modeling the astronomical impact of large constellations.

# References

## Reports discussed during the workshop

## Additional citations, including reports referenced during the workshop

## Acknowledgements

We gratefully acknowledge the cheerful help of Brittany Hong, Monda Korich, and Bill Tuck during the workshop, and the help of Norris Bach, Lourdes Gomez, Valerie Hoag, Jeremy Phillips, Bill Tuck, and Daniel Wang during the organizational phase. This report would not have been possible without them. We especially thank Pablo Trefftz-Posada and Forrest Fankhauser for giving a satellite industry perspective during the workshop. We thank Enrique Allona at Eutelsat for a useful conversation. Finally, we thank the external reviewer Pat Seitzer. The workshop was funded via grant AST-2515440 from the National Science Foundation.